\documentclass[prc,aps,superscriptaddress,floatfix,twocolumn,nofootinbib]{revtex4-2}
\usepackage{graphicx}
\usepackage{makecell}
\usepackage{bm}
\usepackage{color}
\usepackage{ulem}
\usepackage{tikz}

\begin{document}

\title{Direct {\itshape ab initio} calculation of the $^{4}$He nuclear electric dipole polarizability}
\author{Peng Yin}
\affiliation{College of Physics and Engineering, Henan University of Science and Technology, Luoyang 471023, China}
\affiliation{CAS Key Laboratory of High Precision Nuclear Spectroscopy, Institute of Modern Physics, Chinese Academy of
Sciences, Lanzhou 730000, China}
\affiliation{Department of Physics and Astronomy, Iowa State University, Ames, IA 50011, USA}

\author{Andrey~M.~Shirokov}
\affiliation{Skobeltsyn Institute of Nuclear Physics, Lomonosov Moscow State University, Moscow 119991, Russia}

\author{Pieter Maris}
\affiliation{Department of Physics and Astronomy, Iowa State University, Ames, IA 50011, USA}

\author{Patrick J.~Fasano}
\altaffiliation[Present address: ]{Physics Division, Argonne National Laboratory, Argonne, Illinois 60439-4801, USA}
\affiliation{Department of Physics and Astronomy, University of Notre Dame, Notre Dame, Indiana 46556-5670, USA}

\author{Mark A. Caprio}
\affiliation{Department of Physics and Astronomy, University of Notre Dame, Notre Dame, Indiana 46556-5670, USA}

\author{He Li}
\affiliation{CAS Key Laboratory of High Precision Nuclear Spectroscopy, Institute of Modern Physics, Chinese Academy of
Sciences, Lanzhou 730000, China}
\affiliation{School of Nuclear Science and Technology, University of Chinese Academy of Sciences, Beijing 100049, China}

\author{Wei Zuo}
\affiliation{CAS Key Laboratory of High Precision Nuclear Spectroscopy, Institute of Modern Physics, Chinese Academy of
Sciences, Lanzhou 730000, China}
\affiliation{School of Nuclear Science and Technology, University of Chinese Academy of Sciences, Beijing 100049, China}

\author{James P. Vary}
\affiliation{Department of Physics and Astronomy, Iowa State University, Ames, IA 50011, USA}

\begin{abstract}
The calculation of nuclear electromagnetic sum rules by directly solving for numerous eigenstates in a large basis is numerically challenging and has not been performed for $A>2$ nuclei. With the significant progress of high performance computing, we show that calculating sum rules using numerous discretized continuum states obtained with the \textit{ab initio} no-core shell model in the harmonic oscillator basis is achievable numerically. Specifically, we calculate the $^{4}$He electric dipole ($E1$) polarizability,
that is an inverse energy weighted sum rule, employing the Daejeon16 $NN$ interaction. We demonstrate that the calculations are numerically tractable as the dimension of the basis increases and are convergent. Our results for the $^{4}$He electric dipole polarizability are consistent with the most recent experimental data and are compared with those of other theoretical studies employing different techniques and various interactions.
\end{abstract}

 \maketitle

\section{Introduction}
Electromagnetic transitions in atomic nuclei can reveal important information about the dynamical structure of the nucleus itself~\cite{Bacca:2014tla}. Due to the perturbative essence of the electromagnetic interaction, calculations of these observables can be compared in a straightforward way to experimental data, and important features of the strongly interacting nuclear many-body system can be studied. Considering the transitions from the ground state to the low-lying and highly excited states, one can study the sum rules, which can be compared to experiment as well. A sum rule is often associated with spectral integration over a nuclear response function with an energy-dependent weight function, which is related to a reaction cross section of a nucleus due to an external probe. Therefore investigations of nuclear sum rules may provide important information on related reactions. A prominent example, the electric dipole ($E1$) polarizability of a nucleus, which is the inverse energy weighted sum rule of the $E1$ transition and represents the response of the nucleus to two successive electric impulses, is crucial for nuclear photoabsorption reactions~\cite{Friar:1983zza}, Coulomb breakup reactions~\cite{Rodning:1982zz} and astrophysics~\cite{Hagen:2015yea}.

The straightforward calculation of electromagnetic sum rules requires an integral over the continuum states which is computationally challenging. In practice, one often approximates the continuum states by expanding the nuclear wave functions with a complete discrete set of localized basis states, which are then truncated to a finite basis. By solving for the eigenvalue problem in this basis, one can obtain a set of eigenstates which can be regarded as a discretized approximation of the continuum. Under this assumption, the sum rule becomes a sum over the transition probabilities from the ground state to the discretized continuum states. The sum rule is expected to converge to the continuum value as the basis size increases. This expectation was numerically verified only in the case of the deuteron where the diagonalization of the Hamiltonian was used to calculate sum rules~\cite{Bacca:2014tla,Friar:1997fp,Yin:2022zii,Hernandez:2014pwa,Hernandez:2017mof}. The calculations have not previously been extended to $A>2$ nuclei due to the computational costs of solving for numerous eigenstates.

Using alternative techniques, such as the Lorentz (or Stieltjes) integral transform~\cite{Efros:1994iq,Efros:2007nq,Gazit:2006ey,Leidemann:2003ey,Ji:2013oba,Bacca:2014rta,Bacca:2013dma,Miorelli:2016qbk,Bonaiti:2021kkp,Hagen:2015yea,Fearick:2023lyz}  and the Lanczos sum rule method~\cite{Haxton:2005ci,NevoDinur:2014ngu,Stetcu:2008vt,Stetcu:2009py,Baker:2020rbq,Schuster:2013sda}, which avoid solving for numerous eigenstates, has been the only viable way for calculating the sum rules in \textit{ab initio}
approaches up to now. Using these alternative techniques, sum rules have been successfully calculated with \textit{ab initio} approaches, such as hyperspherical harmonics (HH) (for $^4$He)~\cite{NevoDinur:2014ngu,Gazit:2006ey,Leidemann:2003ey,Ji:2013oba}, the coupled-cluster (CC) method (for $^{4,8}$He, $^{16,22}$O and $^{40,48}$Ca)~\cite{Bacca:2014rta,Bacca:2013dma,Miorelli:2016qbk,Bonaiti:2021kkp,Acharya:2022drl,Fearick:2023lyz,Hagen:2015yea}, no-core shell model (NCSM) (for $A\le4$)~\cite{Stetcu:2008vt,Stetcu:2009py,Schuster:2013sda,Baker:2020rbq} and symmetry-adapted no-core shell model (SA-NCSM) (for $^4$He, $^{16}$O, $^{20}$Ne and $^{40}$Ca)~\cite{Baker:2020rbq,Burrows:2023ugy}. Calculations of the sum rules from realistic interactions can also be obtained by equations of motion phonon method~\cite{DeGregorio:2022anr}, but presently is limited by the basis truncation and the truncation of the multiphonon space.

In this letter, we demonstrate that evaluating sum rules by directly solving for numerous eigenstates in a large basis space is feasible for $A>2$ nuclei with the continued major advances in high performance computing. Specifically, we calculate the $^{4}$He $E1$ polarizability by solving for its eigenstates with the \textit{ab initio} NCSM in the harmonic oscillator (HO) basis and achieving numerically tractable results as the basis size increases. The $^{4}$He $E1$ polarizability calculated with various \textit{ab initio} approaches employing different interactions shows significant variations and the available experimental data have not provided strong constraints due to large experimental uncertainties (see below for details). Our calculation with the Daejeon16 nucleon-nucleon ({\it NN}) interaction~\cite{Shirokov:2016ead} is therefore important for both theoretical and experimental studies of the $^{4}$He $E1$ polarizability in the future. This interaction is based on the Entem-Machleidt N$^3$LO chiral effective field theory interaction~\cite{Entem:2003ft}, softened via a similarity renormalization group (SRG) transformation so
as to improve convergence of \textit{ab initio} studies, modified off-shell to mimic three-nucleon forces and provide one of the best descriptions of nuclei with $A\le16$~\cite{Maris:2019etr}. Our detailed study of the convergence properties of $^4$He $E1$ polarizability provides a guide for potential applications to heavier nuclei. Our method can be extended straightforwardly to the calculations of sum rules involving other operators.

In the next Section, we present the theoretical framework adopted in this work. We show the main results in Sec.~\ref{sec:results}. Finally we give a summary of our conclusions and an outlook in Sec.~\ref{sec:conclusions}.

\section{Theoretical methods}
\label{sec:theoryReview}

The $E1$ polarizability of a nucleus, $\alpha_E$, is defined as~\cite{Alder:1956im}
\begin{equation}
\alpha_E=\frac{8\pi}{9}\sum_k\frac{B(E1;J_0\rightarrow J_k)}{E_k-E_0},
\label{eq:pol}
\end{equation}
where $E_0$ and $E_k$ are the energies of the ground and excited states, respectively. $B(E1;J_0\rightarrow J_k)=\sum_{M_k\mu}|\langle J_0M_0 | \hat{D}_{1\mu} | J_kM_k \rangle|^2=|\langle J_0 || \hat{D}_1 || J_k \rangle|^2/(2J_0+1)$ represents the reduced $E1$ transition probability with $E1$ operator $\hat{D}_{1\mu}=\sum_i^A e_i r_i Y_{1\mu}(\bm {r}_i)$. $J_0$ ($J_k$) and $M_0$ ($M_k$) are the ground (excited) state total angular momentum and its projection respectively.

We use the \textit{ab initio} NCSM~\cite{Barrett:2013nh} to calculate the nuclear energy spectrum and the wave functions involved in Eq.~(\ref{eq:pol}). The NCSM has been extensively used recently in studies of $s$- and $p$-shell nuclei
(see, e.\,g., Refs.~\cite{Maris:2016wrd,Maris:2020qne,LENPIC:2022cyu}). In the NCSM, the nuclear wave functions are obtained by diagonalizing the chosen nuclear Hamiltonian in a truncated Slater determinant HO basis characterized by the basis oscillator parameter $\hbar\Omega$.  
We use the $M$ scheme with conserved
parity $\pi$ and projections of the total angular momentum $M=\sum_{i=1}^A m_i$ and charge (isospin projection) $M_T=\sum_{i=1}^A m_{t_i}$. The truncation of the model space is determined and labeled by the number of excitation quanta, $N_{\rm max}$, which corresponds to the total number of HO quanta relative to the minimum number of quanta required by the Pauli principle. We test convergence by showing calculated quantities vs $N_{\rm max}$, and we report how these quantities approach their asymptotic values as $N_{\rm max}$ increases.

\begin{figure}[t]
\includegraphics[width=\columnwidth]{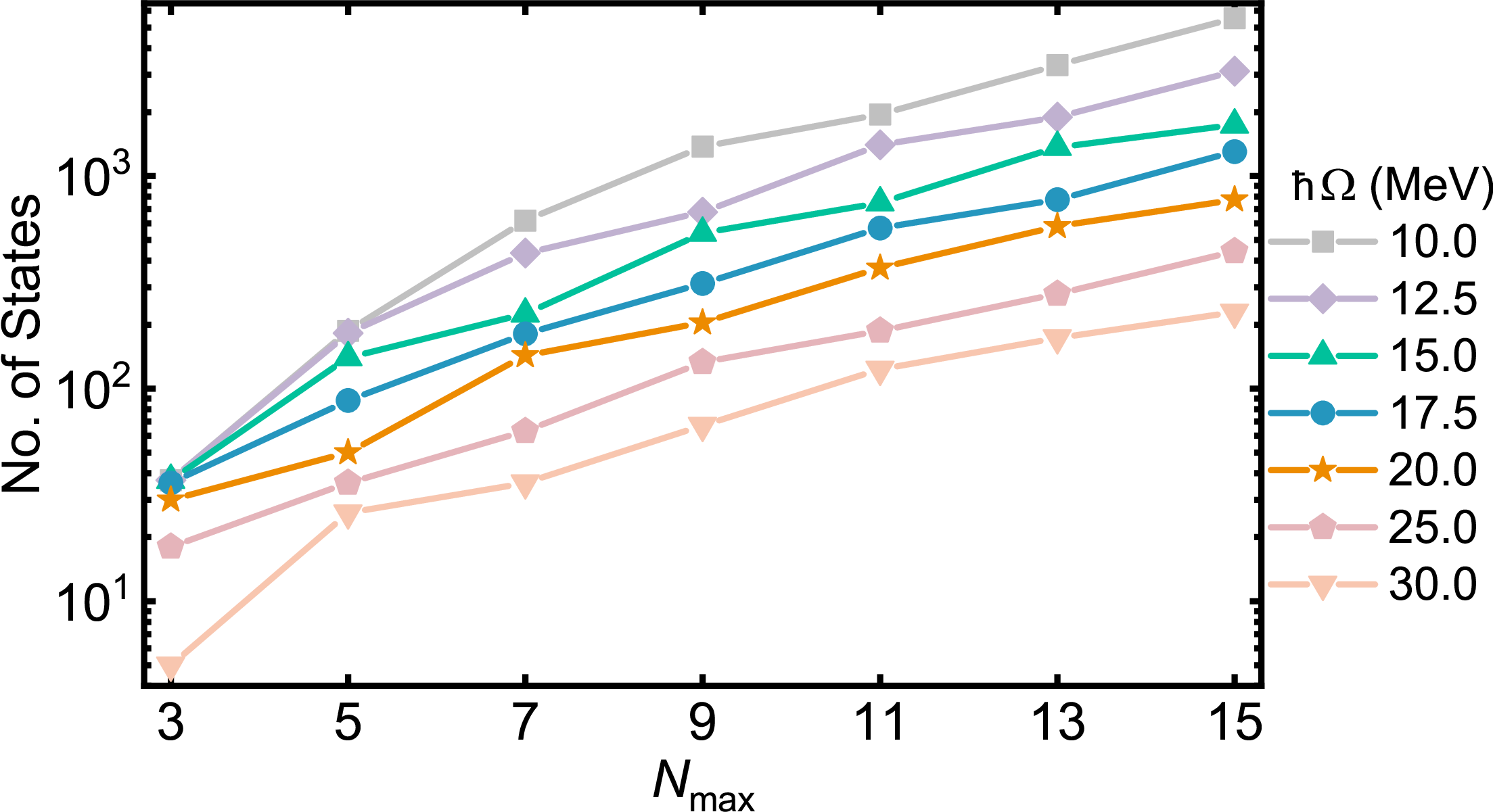}
\caption{(Color online) Number of $1^-$ states in $^4$He up to the excitation energy of $100$ MeV as a function of $N_{\rm max}$ calculated by the NCSM using the Daejeon16 $NN$ interaction with the Coulomb interaction between protons. }\label{fig1}
\end{figure}

In this work, we evaluate the $E1$ polarizability of $^4$He by solving for its eigenstates with the code MFDn using the Lanczos algorithm~\cite{Aktulga:2014im,Maris:2010im,Maris:2022im,Cook:2021im}. We perform the calculations with the Daejeon16 {\it NN} interaction~\cite{Shirokov:2016ead} and add the Coulomb interaction between the protons. The $0^+$ ground state is of normal parity, and is thus obtained from calculations with even $N_{\rm max}$ ($N_{\rm max}=2,4,6,\cdots$). According to the $E1$ selection rules, only $1^-$ excited states are allowed excited states in Eq.~(\ref{eq:pol}). The $1^-$ excited states are of non-normal parity, and are thus obtained from calculations with odd $N_{\rm max}$ ($N_{\rm max}=3,5,7,\cdots$). Within the $M$ scheme, we can obtain the desired $1^-$ states with $M=1$. With a straightforward calculation we would obtain numerous unneeded states with $J\geq2$ at the same time. For example, we obtain $2346$ states (including $1^-$, $2^-$, $\cdots$) below the excitation energy of $100$~MeV
at $N_{\rm max}=11$ with $\hbar\Omega=20$ MeV and $M=1$, whereas only $370$ are our desired $1^-$ states. In order to remove these unneeded states and conserve computational resources, we add the following Lagrange multiplier term to the Hamiltonian~\cite{Whitehead:1977qp}
\begin{equation}
H_{J^2}=\lambda\left(\bm{J}^2-2\right),
\label{eq:HJ}
\end{equation}
with the total angular momentum vector given by $\bm{J}=\sum_{i=1}^A \bm{j}_i$. For example, the $2^-$ ($3^-$) states are shifted upwards in energy by this term by $100$ ($250$)~MeV if we set $\lambda=25$~MeV.
We show in Fig.~\ref{fig1} that the number of $1^-$ states below $100$ MeV excitation energy increases rapidly with $N_{\rm max}$, especially at low $\hbar\Omega$ values. Therefore one of the challenges in calculating the $E1$ polarizability of $^4$He is obtaining necessary $1^-$ states. We employ the Lanczos method with a sufficient number of Lanczos iterations to converge all $1^-$ states up to $100$ MeV of excitation. We then calculate the $E1$ transition matrix elements from the ground state to these $1^-$ states and the resulting $E1$ polarizability.

We adopt the Lawson method~\cite{Whitehead:1977qp,Gloeckner:1974sst} to ensure that only states which are free from spurious center of mass (CoM) excitation, that is, which share the same $0s$ CoM wave function, remain in the calculated spectrum. The CoM wave function does not contribute to the $E1$ transition matrix elements due to the angular momentum and parity selection rules on the CoM degree of freedom, as shown in Ref.~\cite{Caprio:2020heu}.

One should note that the continuum states obtained with the NCSM in the above manner are viewed as a discretized approximation of the continuum. These discretized states become more dense in the continuum as the basis size increases~\cite{NevoDinur:2014ngu}.

\section{Results and discussion}
\label{sec:results}

\begin{figure*}[t]
\includegraphics[width=0.7\linewidth,scale=1.00]{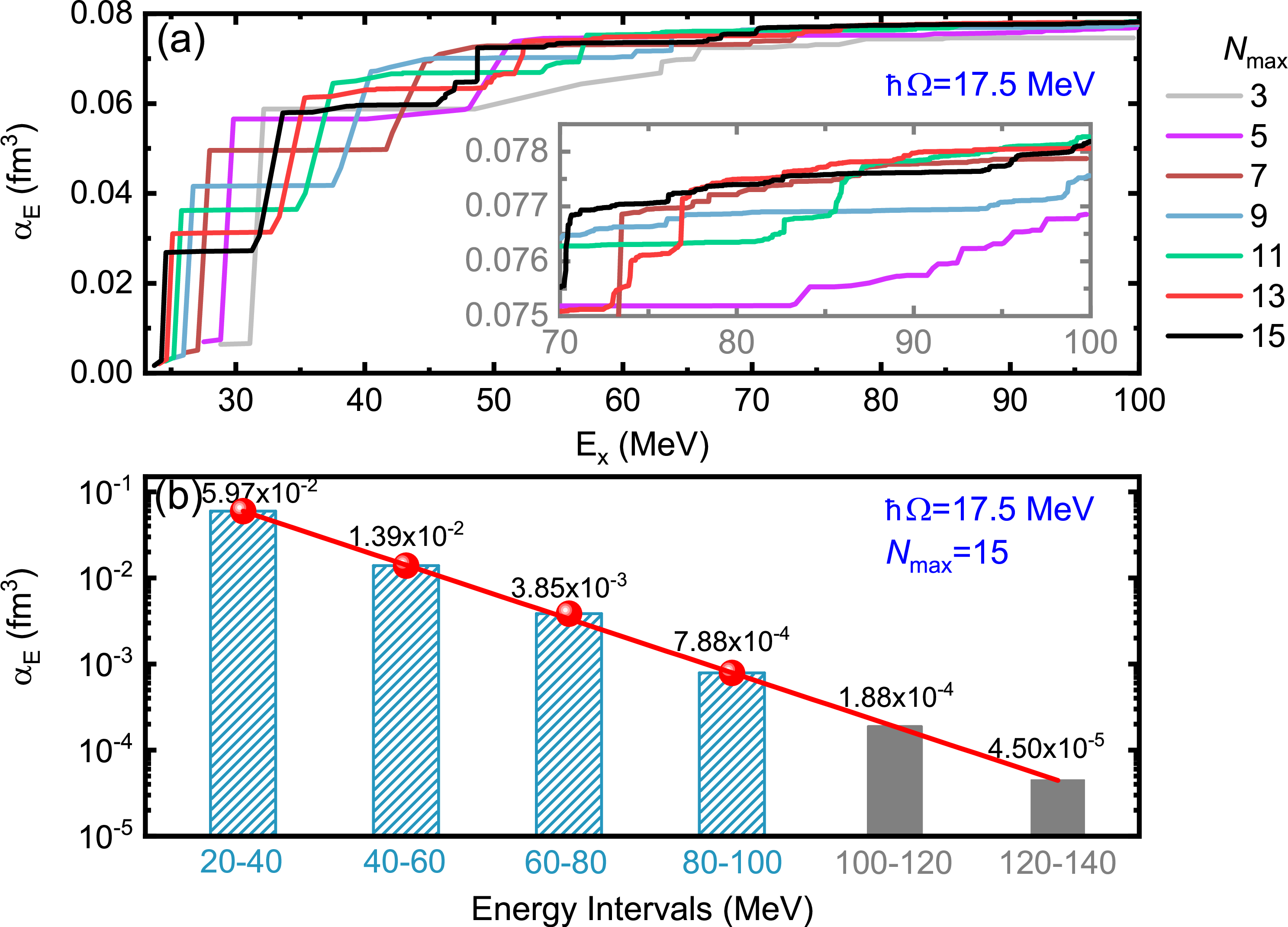}
\caption{(Color online) (a) $E1$ polarizability of $^4$He as a function of the excitation energy $E_x$ calculated within the NCSM at $N_{\rm max}=3{-}15$ with $\hbar\Omega=17.5$ MeV. The inset shows the expanded view of the energy interval from $70$ to $100$ MeV for the $N_{\rm max}\ge 5$ results. (b)~Contributions of different energy intervals to the $E1$ polarizability of $^4$He at $N_{\rm max}=15$ with $\hbar\Omega=17.5$~MeV. Bins filled with hatch lines represent the results calculated with the NCSM and the four solid dots indicate the amplitude of each bin. The solid curve is obtained by exponential fitting to the four solid dots. The bins without hatch lines represent the contributions of $100-120$  and $120-140$ MeV intervals extrapolated with the solid curve.}\label{fig2}
\end{figure*}
In Fig.~\ref{fig2} (a) we present the $E1$ polarizability, $\alpha_E$, of $^4$He as a function of the cutoff in the excitation energy (i.e., the running sum) for various $N_{\rm max}$ with the same $\hbar\Omega=17.5$ MeV. We present in our figures the results marked by $N_{\rm max}$ which is used to calculate the excited $1^-$ states while the ground state is calculated with $N_{\rm max}-1$. We find in Fig.~\ref{fig2} (a) that $\alpha_E$ at first increases rapidly with energy cutoff, for low cutoff energies. However, the growth of $\alpha_E$ slows drastically above roughly $75$ MeV, in particular, at $N_{\rm max}\ge 7$.
The rather small differences of the results for $N_{\rm max}=11{-}15$ at $E_x=100$ MeV indicate the approximate convergence of the NCSM calculations with respect to the basis truncation.

We show in Fig.~\ref{fig2} (b) the calculated contributions of different energy intervals below $100$ MeV excitation energy to the $E1$ polarizability of $^4$He (bins filled with hatch lines). We observe an overall decrease of the contributions of the $1^-$ states with the increase of the excitation energy. The bin of $80-100$ MeV contributes less than $10^{-3}$ fm$^3$. The computational cost depends strongly on the number of Lanczos iterations which determines the truncation to the excitation energy. It becomes computationally prohibitive to account for excitation energies above some point dictated by available computational resources. We therefore truncate the excitation energy at $100$~MeV in the following calculations and estimate the uncertainty caused by this truncation.

\begin{figure}[t]
\includegraphics[width=\columnwidth]{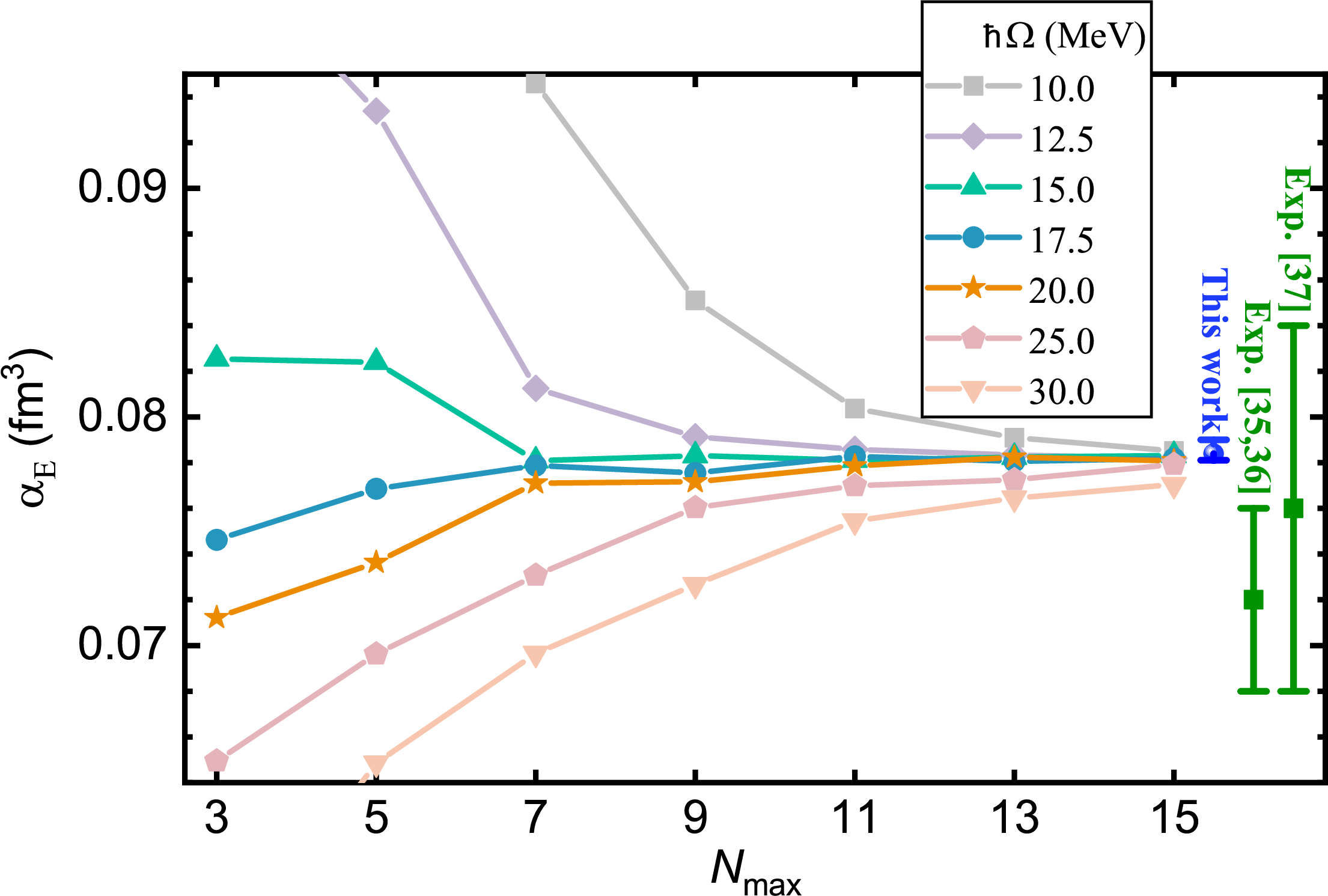}
\caption{(Color online) Electric dipole polarizability of $^4$He as a function of $N_{\rm max}$ for various $\hbar\Omega$ values. The solid dot with an error bar represents the converged result with an estimated uncertainty. Two experimental data along with their quoted uncertainties (solid squares) from Refs.~\cite{Friar:1977cf,Arkatov:1974,Pachucki:2006pq} are shown for comparison.}\label{fig3}
\end{figure}
In Fig.~\ref{fig3} we show the calculated $E1$ polarizability of $^4$He as a function of the truncation parameter $N_{\rm max}$ for
various $\hbar\Omega$ values. The number of the $1^-$ states increases sharply with
decreasing $\hbar\Omega$ at the same $N_{\rm max}$ as shown in Fig.~\ref{fig1}. In order to limit the computational costs, we restrict the NCSM calculations up to $N_{\rm max}=15$ in this work. We obtain all the results in Fig.~\ref{fig3} by retaining $1^-$ states up to $100$ MeV excitation energy.

We observe in Fig.~\ref{fig3} that, as $N_{\max}$ increases, different
basis oscillator parameters $\hbar\Omega$ result in different convergence patterns which are especially visible for small $N_{\rm max}$ values. We find that the results with moderate $\hbar\Omega=15{-}20$ MeV in Fig.~\ref{fig3} show the most rapid convergence at small $N_{\rm max}$. The results for all $\hbar\Omega$ values tend to the same asymptotic value as $N_{\rm max}$ increases. In particular, the results with $\hbar\Omega=15$ and $17.5$~MeV show apparent convergence with some small oscillations, which indicates that calculating the $E1$ polarizability by directly solving for nuclear eigenstates in a large HO basis is numerically achievable.

We use the result at $N_{\rm max}=15$ with $\hbar\Omega=17.5$ MeV, i.\,e., $\alpha_E=0.0782$ fm$^3$, as our prediction to the $E1$ polarizability of $^4$He considering the contribution of the energy interval $0{-}100$ MeV. The uncertainty for calculating nuclear $E1$ polarizability by direct Hamiltonian diagonalization is restricted mainly by two aspects, i.\,e., the truncation of the excitation energy and limited basis size.

In order to estimate the uncertainty stemming from the excitation energy truncation, we fit the amplitudes (represented by four solid dots) of the four bins filled with hatch lines in Fig.~\ref{fig2} (b) with an exponential function, and extrapolate the fitting results up to the two intervals $100-120$ and $120-140$ MeV (bins without hatch lines). Since the fitting results (solid curve in Fig.~\ref{fig2} (b)) coincide well with the four solid dots, we could estimate the uncertainty induced by the excitation energy truncation with the the extrapolated results. We approximate this uncertainty with the extrapolated contributions above $100$ MeV, i.e., $0.0003$ fm$^{-3}$. One should note that the quality of the exponential fit depends on the $N_{\rm max}$ and $\hbar\Omega$ values, as well as on the bin size; the above extrapolation is evaluated only for $N_{\rm max}=15$ with $\hbar\Omega=17.5$ MeV with the bin size of $20$ MeV as shown in Fig.~\ref{fig2} (b).

We estimate the uncertainty due to basis truncation with the difference of the results for $\hbar\Omega=17.5$ and $25$~MeV at the same $N_{\rm max}=15$ in Fig.~\ref{fig3}, i.\,e., $0.0003$~fm$^3$. The $E1$ polarizability of $^4$He obtained with the Daejeon16 interaction in this work is therefore $0.0782^{+0.0006}_{-0.0003}$~fm$^3$. Since the uncertainty due to the energy truncation can only be positive, the upper error bar is obtained simply with the sum of the above two uncertainties while the lower error bar is contributed only by the basis truncation. We show our converged result with the estimated uncertainty in Fig.~\ref{fig3}. All the calculated results at $N_{\rm max}=15$ fall into our estimated uncertainty region with the only exception of the result for $\hbar\Omega=30$ MeV, which indicates that we have achieved good convergence and our estimation to the theoretical uncertainty is reasonable. We observe in Fig.~\ref{fig3} that our predicted $E1$ polarizability of $^4$He is consistent with the most recent experimental data~\cite{Pachucki:2006pq}.


\begin{table}[t]
\vspace{-2ex}
\renewcommand\arraystretch{1.5}
\caption{\label{tab:table1}$^4$He $E1$ polarizability. Our result in comparison with experimental data and those obtained by various
{\itshape ab initio} approaches employing different techniques and various internucleon interaction models.}
\begin{tabular*}{\linewidth}{l l l}
  \hline\hline
  Theo./Exp. & Interaction  & $\alpha_E$ (fm$^3)$
  \\ \hline
  \thead[l]{NCSM\\ $[$This work$]$} & {\it NN} (Daejeon16) & $0.0782^{+0.0006}_{-0.0003}$
  \\ \hline
  HH~\cite{Gazit:2006ey} & {\it NN} (AV18) + 3{\it N} (UIX) & $0.0655(4)$ \\
  HH~\cite{Leidemann:2003ey} & {\it NN} (MT-\uppercase\expandafter{\romannumeral1}/\uppercase\expandafter{\romannumeral3}) & $0.076$  \\
  HH~\cite{Ji:2013oba}      & [{\it NN} (N$^3$LO) + 3{\it N} (N$^2$LO)]$_{\rm OLS}$ & $0.0694$  \\
  NCSM~\cite{Stetcu:2009py}  & [{\it NN} (N$^3$LO) + 3{\it N} (N$^2$LO)]$_{\rm OLS}$ & $0.0683(14)$ \\
  NCSM~\cite{Schuster:2013sda}    & \thead[l]{$[${\it NN} (N$^3$LO) + 3{\it N} (N$^2$LO)$]$$_{1.8}$\\$[${\it NN} (N$^3$LO) + 3{\it N} (N$^2$LO)$]$$_{3.0}$} & \thead[l]{0.07093(5)\\ 0.06861(5)}  \\
  NCSM~\cite{Baker:2020rbq}    & {\it NN} (N$^3$LO) & $0.084(3)$  \\
  SA-NCSM~\cite{Baker:2020rbq}    & {\it NN} (N$^3$LO) & $0.077(3)$  \\
  \thead[l]{NCSM\\ SA-NCSM}\cite{Baker:2020rbq}    & {\it NN} (NNLO$_{\rm opt}$) & $0.0680$ \\
  CC~\cite{Miorelli:2016qbk}& {\it NN} + 3{\it N} (NNLO$_{\rm sat}$) & $0.0735(1)$  \\ \hline
  Exp.~\cite{Friar:1977cf,Arkatov:1974}    & - & $0.072(4)$  \\
  Exp.~\cite{Pachucki:2006pq}    & - & $0.076(8)$  \\
  \hline\hline
\end{tabular*}
\end{table}

We present our results in Table~\ref{tab:table1}
and compare with some alternative {\it ab initio} calculations of the $^4$He $E1$ polarizability. We also present two experimental data along with their quoted uncertainties from Refs.~\cite{Friar:1977cf,Arkatov:1974,Pachucki:2006pq} for comparison. The {\it ab initio} calculations we quote in Table~\ref{tab:table1} for comparison use the following {\it NN} and three-nucleon (3{\it N}) interactions:
 {\it NN} (AV18)~\cite{Wiringa:1994wb}, 3{\it N} (UIX)\cite{Pudliner:1997ck}, {\it NN} (MT-\uppercase\expandafter{\romannumeral1}/\uppercase\expandafter{\romannumeral3})~\cite{Malfliet:1968tj},
 {\it NN} (N$^3$LO)~\cite{Entem:2003ft}, 3{\it N} (N$^2$LO)~\cite{Navratil:2007zn}, {\it NN} (NNLO$_{\rm opt}$)~\cite{Ekstrom:2013kea} and {\it NN}+3{\it N} (NNLO$_{\rm sat}$)~\cite{Ekstrom:2015rta}. The subscript OLS denotes that the results in Refs.~\cite{Ji:2013oba,Stetcu:2009py} are obtained with a Okubo-Lee-Suzuki renormalization~\cite{Okubo:1954zz,Suzuki:1980yp,Suzuki:1982yp} of internucleon interactions.
A SRG-evolved {\it NN}+3{\it N} Hamiltonian and a self-consistent SRG-evolved $E1$ operator are used in Ref.~\cite{Schuster:2013sda}. The subscripts $1.8$ and $3.0$ in Table~\ref{tab:table1} denote the SRG-evolved scale parameters in fm~\cite{Schuster:2013sda}. The NCSM approach was also used in Refs.~\cite{Stetcu:2009py,Schuster:2013sda,Baker:2020rbq} to calculate the $^4$He polarizability. Besides the different Hamiltonians adopted, one of the major differences between our calculation and Refs.~\cite{Stetcu:2009py,Schuster:2013sda,Baker:2020rbq} is that we solve for numerous eigenstates which is avoided by Refs.~\cite{Stetcu:2009py,Schuster:2013sda,Baker:2020rbq}.
We can see in Table~\ref{tab:table1} significant differences for
$^4$He $E1$ polarizabilities calculated with various internucleon interaction models, ranging from $0.0655(4)$ to $0.084(3)$ fm$^3$.
From the results shown in Table~\ref{tab:table1}, we infer that the nuclear $E1$ polarizability may provide important constraints on nuclear interactions. These results provide motivation for further improvements of both theoretical calculations and experimental measurements, such as the photoabsorption reactions and the Coulomb breakup reactions.

\vspace{1mm}
\section{conclusions and outlook}
\label{sec:conclusions}

In conclusion, we calculated the electric dipole ($E1$) polarizability of $^4$He through directly solving for its numerous eigenstates by means of the \textit{ab initio} NCSM in the HO basis. We performed calculations with the realistic Daejeon16
$NN$ interaction. We considered the $E1$ transitions from the ground state to all calculated $1^-$ states below an excitation energy of $100~$MeV. In order to retain only $1^-$ states and exclude the unneeded states with higher angular momentum, we added an angular momentum Lagrange multiplier term to the Hamiltonian. The running sum exhibits a good convergence pattern as the cutoff of excitation energy increases. The $E1$ polarizability of $^4$He converges to the third significant figure as the basis size increases. Our predicted $E1$ polarizability of $^4$He, $0.0782^{+0.0006}_{-0.0003}$ fm$^3$, is consistent with the most recent
experimental data. We expect that direct calculations may also be applicable to other sum rules in nuclei. The results obtained with this method can also be used to benchmark evaluations of nuclear sum rules with other techniques.

Our study provides motivation for similar calculations in heavier nuclei. For instance, the {\itshape ab initio} calculations of the $E1$ polarizability of the two-neutron halo nucleus $^6$He, may strongly constrain the measurement of its $E1$ response function which disagrees significantly among different experiments~\cite{Aumann:1999mb,Wang:2002sm,Sun:2021efo}. Although the theoretical uncertainties of calculations in heavier nuclei will be larger, the results with estimated uncertainties could nevertheless provide useful predictions for experiment and theoretical comparisons.

\section*{Acknowledgments}
We acknowledge helpful discussions with Chen Ji and Xingbo Zhao.
A portion of the computational resources are provided by the National Energy Research Scientific Computing Center (NERSC), a U.S. Department of Energy Office of Science User Facility located at Lawrence Berkeley National Laboratory, operated under Contract No. DE-AC02-05CH11231 using NERSC award NP-ERCAP0020944, NP-ERCAP0023866 and NP-ERCAP0028672.
Peng Yin and Wei Zuo are supported by the National Natural Science Foundation of China (Grant
Nos.~11975282, 11705240, 11435014), the Strategic Priority Research Program of Chinese Academy of Sciences, Grant No.~XDB34000000, the Key Research Program of the Chinese Academy of Sciences under Grant No.~XDPB15, and the Natural Science Foundation of Gansu Province under Grant No.~23JRRA675. A.~M.~Shirokov is thankful to the Chinese Academy of Sciences President's International Fellowship Initiative Program (Grant No.~2023VMA0013) which supported his visits to Lanzhou where a part of this work was performed and acknowledges the hospitality of Chinese colleagues during these visits. This material is based upon work supported by the U.S.~Department of
Energy, Office of Science, under Award Nos.~DE-FG02-95ER40934 (M.A.C.),
DE-SC0023495 (P.M./J.P.V.), and DE-SC0023692 (J.P.V.).

\end{document}